\documentclass[aps,prl,twocolumn,superscriptaddress,showpacs,floatfix,nobibnotes]{revtex4}

\usepackage{epsfig}
\usepackage{epstopdf}
\usepackage{color}

\usepackage{graphicx}
\usepackage{longtable}
\usepackage{CJK}

\usepackage{graphicx}
\usepackage{longtable}
\usepackage{CJK}

\begin{document}

\title{The Heine-Stieltjes correspondence and the\\
 polynomial approach to the standard pairing problem}

\begin{CJK*}{GBK}{song}

\author{Feng Pan}
\affiliation{Department of Physics, Liaoning Normal University,
Dalian 116029, China}\affiliation{Department of Physics and
Astronomy, Louisiana State University, Baton Rouge, LA 70803-4001,
USA}

\author{Xin Guan}
\affiliation{Department of Physics, Liaoning Normal University,
Dalian 116029, China}

\author{Mingxia Xie}
\affiliation{Department of Physics, Liaoning Normal University,
Dalian 116029, China}

\author{ Lina Bao}
\affiliation{Department of Physics, Liaoning Normal University,
Dalian 116029, China}

\author{J. P. Draayer}
\affiliation{Department of Physics and Astronomy, Louisiana State
University, Baton Rouge, LA 70803-4001, USA}
\date{\today}

\begin{abstract}
A new approach for solving the Bethe ansatz (Gaudin-Richardson) equations of
the standard pairing problem is established based on the Heine-Stieltjes correspondence.
For $k$ pairs of valence nucleons on $n$ different single-particle
levels, it is found that
solutions of the Bethe ansatz equations can be obtained from
one $(k+1)\times (k+1)$ and one $(n-1)\times (k+1)$ matrices, which
are associated with the extended Heine-Stieltjes and Van Vleck polynomials,
respectively. Since the coefficients in these polynomials
are free from divergence with variations in contrast
to the original Bethe ansatz equations,
the approach thus provides with a new efficient and systematic way to solve
the problem, which, by extension, can also be used
to solve a large class of Gaudin-type quantum many-body problems
and to establish a new efficient angular momentum projection method
for multi-particle systems.
\end{abstract}

%
%

\maketitle

\end{CJK*}

\newpage

\parindent=20pt

It is well known that the pairing force, similar to that
in the Bardeen-Cooper-Schrieffer (BCS) theory of superconductors~\cite{[1]},
as one of main residual interactions introduced to
the nuclear shell model, is key to manifest
ground state properties and low energy spectroscopy of nuclei,
such as binding energies, odd-even effects,
single-particle occupancies, excitation spectra, electromagnetic
transition rates, beta-decay probabilities, transfer reaction
amplitudes, low-lying collective modes, level densities, and moments
of inertia, and so on~\cite{[2]}-\cite{[4]}.
Unlike electrons in solids,
the drawbacks of the application of the BCS theory and its extensions
to nuclei are noticeable due to the fact that the number of valence nucleons
under the influence of the pairing force is too few
to be treated by such particle-number nonconservation (quasi-particle)
approximations~\cite{[5]}-\cite{[6]}.

Exact solutions to the standard pairing problem was
first obtained by Richardson, now referred to as the
Richardson-Gaudin method~\cite{[7]}-\cite{[8]}.
Recently, extensions to the Richardson-Gaudin theory have also
been made by using the Bethe ansatz methodology~\cite{[9]}-\cite{[12]}.
The advantage of the Richardson-Gaudin solutions lies in
the fact that the huge matrix in the Fock subspace
is reduced to a set of equations, of which the number
equals exactly to that of pairs of valence particles
involved. However, less attention had been paid to the Richardson's solutions
of the pairing problem in realistic calculations
mainly because the non-linear Bethe ansatz
(Gaudin-Richardson) equations (BAEs) involved are very difficult to be solved
numerically, especially for large size systems. Though there were a number of
authors showing their efforts in designing algorithms for solutions
with promising results~\cite{[13]}-\cite{[18]},
obviously efficient procedure for solving the problem
seems still unclear.
Thus, a simple and clear approach to the problem is in demand.

The Hamiltonian of the standard pairing model is given by
\begin{equation}
\hat{H}=\sum_{j=1}^{n}\epsilon
_{j}\hat{n}_{j}-G\sum_{jj^{\prime}}S_{j}^{+}S^{-}_{j^{\prime}},
  \label{eqno1}
  \end{equation}
where $n$ is the total number of levels considered, $G>0$ is the
overall pairing strength, $\{\epsilon _{j}\}$ are unequal single-particle
energies, $\hat{n}_{j}=\sum_{m} a_{jm}^{\dagger }a_{jm}$ is
the number operator for valence particles in the $j$-th level,
and $S_{j}^{+}=\sum_{m}(-)^{j-m} a_{jm}^{\dagger }a_{j~-m}^{\dagger }$
($S^{-}_{j}=(S_{j}^{+})^{\dagger}$)
are pair creation (annihilation) operators.
Since the formalism for even-odd systems is similar, in the
following, we only focus on the even-even seniority zero case.
According to the Richardson-Gaudin method,
$k$-pair eigenstates of (1) can be written as

\begin{equation}
\vert k;\zeta\rangle=S^{+}(x_{1})S^{+}(x_{2})\cdots
S^{+}(x_{k})\vert 0\rangle,
\end{equation}
where  $\vert 0\rangle$ is  the pairing vacuum state satisfying
$S^{-}_{j}\vert 0\rangle=0$ for all $j$,
$x_{i}$ ($i=1,2,\cdots,k$)
are spectral parameters to be determined.
It can then be verified by using the corresponding
eigen-equation that (2) is the eigenstates of (1) only when the
spectral parameters $x_{i}$ ($i=1,2,\cdots, k$)
satisfy the following set of BAEs:

\begin{equation}\label{3}
1-2G\sum_{j}{\rho_{j}\over{x_{i}-2\epsilon_{j}}}
-2G\sum_{j(\neq i)}{1\over{x_{i}-x_{j}}}
=0,
\end{equation}
where $\rho_{j}=-(j+1/2)/2$, with the corresponding eigen-energy given by
$E_{n,~k}=\sum_{i=1}^{k}{{x_{i}}}$.

Actually, as shown by Heine and Stieltjes,
there is a one-to-one correspondence
between every set of the Gaudin-Richardson type equations (BAEs)
and a set of orthogonal polynomials,
called by us the extended Heine-Stieltjes polynomials. Roots of these BAEs
are zeros of the polynomials,
which can be interpreted as stable equilibrium positions in two
dimensional complex plane for a set of free unit charges
in an external electrostatic field~\cite{[19]}.
The link between Richardson's BCS pairing
model for nuclei and the corresponding electrostatic problem was
thus established~\cite{[20]}.
According to Heine-Stieltjes correspondence, for nonzero
pairing strenght $G$,
the polynomials $y(x)$ with zeros corresponding
to the solutions of Eq. (\ref{3}) should satisfy
the following second-order Fuchsian equation:
\begin{equation}\label{4}
A(x)y^{\prime\prime}(x)+B(x)y^{\prime}(x)-V(x)y(x)=0,
\end{equation}
where $A(x)=\prod_{j=1}^{n}(x-2\epsilon_{j})$
is a polynomial of degree $n$,
$B(x)$ is the polynomial with

\begin{equation}
B(x)/A(x)=\sum^{n}_{j=1}{2\rho_{j}\over{x-2\epsilon_{j}}}-{1\over{G}}\, ,
\end{equation}
and $V(x)$ is called Van Vleck polynomials~\cite{[19]} of degree
$n-1$, which needs to be determined according to Eq. (\ref{4}).
In the original electrostatic analogue considered by Heine and Stieltjes~\cite{[19]},
the parameters $\{\rho_{j}\}$ acting
as fixed charges should all be positive with no external electrostatic field,
$1/G\rightarrow0$. Therefore, the polynomials $y(x)$
satisfying Eq. (\ref{4}) with  negative $\{\rho_{j}\}$  and $1/G\neq0$
are thus called the extended Heine-Stieltjes polynomials, which
tend to be the original Heine-Stieltjes polynomials with negative $\{\rho_{j}\}$
in the $G\rightarrow\infty$ limit.

In search for polynomial solutions of (\ref{4}), we write
\begin{equation}\label{6}
y(x)=\sum_{j=0}^{k}a_{j}x^{j},~~V(x)=\sum_{j=0}^{n-1}b_{j}x^{j},
\end{equation}
where $\{a_{j}\}$ and $\{b_{j}\}$ are the expansion coefficients to be determined.
Substitution of (\ref{6}) into Eq. (\ref{4}) yields two matrix equations,
the condition that coefficients in front of $x^{i}$ ($i=0,\cdots,k$) must be zero
generates a $(k+1)\times(k+1)$ matrix {\bf F} with ${\bf F}{\bf v}=b_{0}{\bf v}$, where
the eigenvector ${\bf v}$ of {\bf F} is just the expansion coefficients
${\bf v}=\{a_{0},\cdots,a_{k}\}$,
while the condition that
coefficients in front of $x_{i}$  ($i=k+1,\cdots,n+k-1$)
must be zero generates another $(n-1)\times (k+1)$  upper-triangular matrix {\bf P}
with ${\bf P}{\bf v}=0$, which provides with
unique solution of $b_{i}$ ($i=1,\cdots,n-1$)
in terms of $\{a_{j}\}$.
Entries of the two matrices are all linear with the coefficients
$\{b_{1},b_{2},\cdots,b_{n-1}\}$.
Matrices {\bf F} and {\bf P}
can easily be constructed, for which a simple MATHEMATICA code
is provided~\cite{[21]}.

Let the single-particle energies
satisfy the interlacing condition $\epsilon_{1}<\cdots<\epsilon_{n}$.
Real parts of zeros of $y(x)$ satisfy
the interlacing condition,
$-\infty<{\bf Re}(x_{1})<{\bf Re}(x_{2})<\cdots<{\bf Re}(x_{k})<+\infty$, where
${\bf Re}(x_{i})$ lies in one of the $n+1$ intervals
$(-\infty,\epsilon_{1})$, $(\epsilon_{1},\epsilon_{2})$,
$\cdots$, $(\epsilon_{n-1},\epsilon_{n})$, and $(\epsilon_{n},+\infty)$. It should be
noted that many ${\bf Re}(x_{i})$ of adjacent zeros
may lie within the same interval. When $G\rightarrow\infty$,
there will be only $n$ intervals with $(-\infty,\epsilon_{1})$
being removed. The number of
different such allowed configurations gives the possible solutions
of $y(x)$ and the corresponding $V(x)$.
The number of solutions of $y(x)$,
excluding those with sum of zeros of $y(x)$ complex, should equal to
the number of levels produced by the standard pairing model, which is
given by

\begin{equation}
\eta(n,k)=\sum_{p_{1}=0}^{-2\rho_{1}}\cdots\sum_{p_{n}=0}^{-2\rho_{n}}\delta_{q,k}\, ,
\end{equation}
where $q=\sum_{i=1}^{n} p_{i}$.
When $\rho_{i}=-1/2$ for any $i$, which corresponds to the case of the Nilsson mean-field plus
pairing model, $\eta(n,k)=n!/((n-k)!k!)$.
Furthermore, if we set $a_{k}=1$ in $y(x)$, the coefficient $a_{k-1}$ must equal to
negative sum of zeros of $y(x)$ with
$a_{k-1}=-E_{n,k}=-\sum_{i=1}^{k}x_{i}$.
Therefore, the solution corresponding to the largest real $a_{k-1}$
is that for the ground state of the system considered;
those corresponding to the next largest real $a_{k-1}$
is that of the first excited state; and so on.
In the standard pairing model, the
solution with the same $a_{k-1}$
is unique except complex conjugation and permutations
within $\{x_{i}\}$,
which will be helpful in simplifying the calculation process,
especially when only a few low-lying states are needed in the
application. Since the coefficients $\{a_{j}\}$ and $\{b_{j}\}$
in {\bf F}, {\bf P}, and {\bf v} are free from
divergence with arbitrary variations in contrast
to the original Bethe ansatz equations (3),
one can use any standard recursive or
iteration method to solve the problem with
arbitrary initial values of these coefficients
as desired.
Because solving the eigen-equation
${\bf F}{\bf v}=b_{0}{\bf v}$, in which ${\bf F}$ is a $(k+1)\times(k+1)$ matrix,
is the only CPU time consuming operation involved,
the CPU time needed in the process should always be reasonable
for $k\sim 10^{1}${--}$10^{3}$ and $n\sim10^{1}${--}$10^{2}$
sufficient to realistic applications in nuclear physics.

To demonstrate the new approach,
we consider a simple example of
$k=5$ pairs in the sixth major shell with $n=5$ levels, $1h_{7/2}$,
$2d_{5/2}$, $2d_{3/2}$, $3s_{1/2}$, and $1h_{11/2}$, which is
related to the application of the solution to Sm isotopes~\cite{[17]},
and difficult to be solved by directly using the BAEs (3).
We set single-particle energies to be equal spacing with $\epsilon_{i}=i$,
and the overall pairing strength $G=0.5$.
This example can now be dealt with easily by using the new polynomial
approach, of which the number of solutions $y(x)$ is $\eta(5,5)=71$.
First $5$ sets of zeros of the corresponding polynomials $y(x)$
and the corresponding coefficient
$a_{k-1}$ are listed in Table I. The results can be obtained
from the MATHEMATICA on a PC with in a minute.
However,
we observed there are some solutions with complex $a_{k-1}$,
and some solutions are very close to each other like degenerate,
which happened when MATHEMATICA built-in functions were used.
Thus, the total number of solutions obtained is
greater than $\eta(n,k)$, of which solutions with complex $a_{k-1}$ should
be discarded.
Because there is little to be known about these polynomials with negative charges,
further study is needed to see whether
there are indeed solutions with complex $a_{k-1}$.
Both complex $a_{k-1}$ and near degenerate issues
may all be due to the Newtonian iteration adopted
in the original MATHEMATICA package for solving
a set of equations. Special codes designed suitable
for the approach seem needed to overcome the ambiguity.

\begin{table}[ht]
\caption{First $5$ sets of zeros of the possible solutions of the extended
Heine-Stieltjes polynomials and the corresponding
eigen-energies (in arbitrary unit)
of the standard pairing model Hamiltonian (1) in the case of
$k=5$ pairs in the sixth major shell with $n=5$ single-particle levels $1h_{7/2}$,
$2d_{5/2}$, $2d_{3/2}$, $3s_{1/2}$, and $1h_{11/2}$.
The single-particle energies used are $\epsilon_{i}=i$,
and the overall pairing strength $G=0.5$.}
\label{energy} \vspace*{-6pt}
\begin{center}
\begin{tabular}{cc}\hline\hline
Zeros of the polynomials &$\sum_{i=1}^{5}x_{i}$ \\ \hline
$x_{1}=-1.4993,~x_{2}=-1.1412 - 2.1396{\imath},$~~~~~~~~~~~ &$-3.6158$\\
$x_{3}=-1.1412 + 2.1396{\imath},~x_4 =0.0829 - 4.5018{\imath},$~\\
$x_5 =0.0829 + 4.5018{\imath}$~~~~~~~~~~~~~~~~~~~~~~~~~~~~~~~~~~~\\
$~~x_{1}=-0.5078 - 1.0411{\imath},~x_{2}=-0.5078 + 1.0411{\imath},$ &$~3.0299$\\
$x_{3}=0.5469 - 3.3066{\imath},~x_4=0.5469 + 3.3066{\imath},$~~~~\\
$x_5=2.9517~~~~~~~~~~$~~~~~~~~~~~~~~~~~~~~~~~~~~~~~~~~~~~~~~&\\
$~~x_1=-0.9234 - 1.0718\imath,~x_{2}=-0.9234+ 1.0718\imath,$&$~3.5444$\\
$x_{3}=0.0573 - 3.3613\imath,~x_4=0.0573 + 3.3613\imath,$~~~~&\\
$x_5=5.2767$~~~~~~~~~~~~~~~~~~~~~~~~~~~~~~~~~~~~~~~~~~~~~~~~&\\
$~~x_1=-1.1244-1.0987\imath,~x_2=-1.1244 + 1.0987\imath,$ &$~4.8379$\\
$~~x_3=-0.1739-3.4422\imath,~x_4=-0.1739 + 3.4422\imath,$\\
$x_5=7.4346$~~~~~~~~~~~~~~~~~~~~~~~~~~~~~~~~~~~~~~~~~~~~~~~~&\\
$~~x_1=-1.2032- 1.1109\imath,~x_2=-1.2032 + 1.1109\imath,$&$~5.77020$\\
$~~x_3=-0.2619 -3.4804\imath,~x_4=-0.2619 + 3.4804\imath,$&\\
$x_5=8.7004$~~~~~~~~~~~~~~~~~~~~~~~~~~~~~~~~~~~~~~~~~~~~~~~~&\\
\hline \hline
\end{tabular}
\end{center}
\end{table}

In addition, as shown in our previous study~\cite{[22]},
a new angular momentum projection method for
multi-particle systems can be established
based on BAEs similar to (3). In fact, for $n$ angular momenta $j_{i}$
($i=1,2,\cdots,n$), the multi-particle state with total
angular momentum $J=\sum_{i}j_{i}-k$ can be written
as

\begin{equation}\label{7}
\vert \eta,~J,~M=J\rangle=J^{-}(x_{1})J^{-}(x_{2})\cdots
J^{-}(x_{k})\vert {\rm h.w.}\rangle,
\end{equation}
where $\eta$ is a quantum number needed to resolve
the multi-occurrence of $J$,
$\vert {\rm h.w.}\rangle$ is the highest weight
single-particle product state with $\vert j_{1},~m_{1}=j_{1},\cdots, j_{n},~m_{n}=j_{n}\rangle$,
and

\begin{equation}
J^{-}(x)=\sum_{i=1}^{n}{1\over{x-2\epsilon_{i}}}J^{-}_{i},
\end{equation}
in which $J^{-}_{i}$ is the angular momentum lowering operator
only acting on the $i$-th single-particle state $\vert j_{i},~m_{i}\rangle$,
and $\epsilon_{i}$ ($i=1,2,\cdots,n$) can be taken as any set of unequal numbers~\cite{[22]}.
Acting the total angular momentum raising operator $J^{+}$ to (\ref{7})
with $J^{+}\vert \eta,~J,~M=J\rangle=0$, one obtains
the BAEs of $\{x_{i}\}$ the same as Eq. (\ref{3}) in the replacement
$\rho_{{i}}=-(j_{i}+1/2)/2$ in the $G\rightarrow\infty$ limit.
Therefore, once the solutions of (3) in the $G\rightarrow\infty$ limit are obtained,
the resultant $\{x_{i}\}$  once and for all
determine the multi-particle state with good angular momentum $J$.
The number of solutions of (3)
equals exactly to the number of occurrence of $J$ for the given system.
The polynomial solutions (4) in this case exactly
become the Heine-Stieltjes polynomials mentioned previously.
This angular momentum projection is certainly much simpler
than the projection operator technique~\cite{[3]}
and that based on the permutation group method~\cite{[bie]}.

In summary, we have established a new approach
for solving the standard pairing problem based on
a sound mathematical foundation~{---}~the extended
Heine-Stieltjes polynomials and
the corresponding Van Vleck polynomials satisfying
the polynomial solutions of the second order
Fuchsian equation. Thus, we reach the go
of the Richardson-Gaudin theory via the Heine-Stieltjes
correspondence, from which
the exact solutions to the problem can practically
be realized based on two matrix equations.
The approach can easily be extended
and applied to solve a large class of Gaudin-type
quantum many-body problems. A new efficient
angular momentum projection method for multi-particle
systems is thus proposed as a byproduct, of which
the application to either boson or fermion systems
will be studied elsewhere.

Support from the U.S. National Science Foundation (PHY-0500291 \& OCI-0904874),
the Southeastern Universities Research Association, the Natural Science
Foundation of China (10775064), the Liaoning Education Department
Fund (2007R28), the Doctoral Program Foundation
of State Education Ministry of China (20102136110002),
and the LSU--LNNU joint research program (9961) is
acknowledged.



\end{document}